\documentclass[11pt,a4paper]{article}
\usepackage[utf8]{inputenc}

\usepackage{amsmath,amssymb}
\usepackage{graphicx}
\usepackage{subfigure,slashed}
\usepackage{color}
\usepackage{multirow}
\usepackage{jcappub}

\usepackage{array}
\usepackage{soul}
\newcolumntype{L}[1]{>{\raggedright\let\newline\\\arraybackslash\hspace{0pt}}m{#1}}
\newcolumntype{C}[1]{>{\centering\let\newline\\\arraybackslash\hspace{0pt}}m{#1}}
\newcolumntype{R}[1]{>{\raggedleft\let\newline\\\arraybackslash\hspace{0pt}}m{#1}}

\newcommand{\be}{\begin{eqnarray*}}
\newcommand{\ee}{\end{eqnarray*}}

\newcommand{\bee}{\begin{eqnarray}}
\newcommand{\eee}{\end{eqnarray}}
\newcommand{\beeq}{\begin{equation}}
\newcommand{\eeeq}{\end{equation}}

\author[a]{Spyros Argyropoulos,}
\author[b]{Clare Burrage,}
\author[c]{and Christoph Englert}

\affiliation[a]{Physikalisches Institut, Universit\"at Freiburg, Hermann-Herder Str.~3a, 79104 Freiburg,\\ Germany}
\affiliation[b]{School of Physics and Astronomy, University of Nottingham, Nottingham, NG7 2RD,\\ United Kingdom}
\affiliation[c]{School of Physics and Astronomy, University of Glasgow, Glasgow, G12 8QQ,\\ United Kingdom}

\emailAdd{Spyros.Argyropoulos@cern.ch}
\emailAdd{clare.burrage@nottingham.ac.uk}
\emailAdd{Christoph.Englert@glasgow.ac.uk}

\arxivnumber{}

\title{Density dependent displaced vertex signatures as a novel probe of light dark sector scalars at the LHC}

\abstract{Dynamical theories of dark energy predict new degrees of freedom with particular environmental sensitivity to
avoid constraints on fifth forces. We show that the similar, yet complementary multi-purpose detector setup of the ATLAS
and CMS experiments provides a unique opportunity to place sensitivity on such scenarios in a narrow, yet relevant
parameter range. Furthermore, our investigation gives rise to a novel phenomenological signature that the LHC experiments
can pursue to exploit their complementary detector design from a BSM perspective.}

\begin{document}

\maketitle

\section{Introduction}
\label{sec:intro}
New physics searches at the energy frontier, chiefly performed at the Large Hadron Collider (LHC) at CERN, have so far not revealed any significant deviation from the established Standard Model (SM) of Particle Physics. Yet, the flaws inherent to the SM suggest that it should only be a good description of microscopic physics up to a few hundred GeV. The experimental developments at the LHC have put a considerable strain on theoretically well-motivated ultraviolet (UV) completions of the SM. This has led to a resurgence of effective field theory methodology on the one hand, and signature-driven phenomenological analyses on the other hand. The combination of both avenues aims to close sensitivity gaps of the LHC experiments that present themselves when moving away from traditional beyond the SM (BSM) scenarios.

Parallel to physics analyses at microscopic scales, a range of experiments has established the $\Lambda$CDM as a successful description of the Universe at cosmological scales and times. Dynamical theories of dark energy (DE) \cite{Copeland:2006wr} provide a theoretical explanation of the continued expansion of the universe and usually employ scalar fields to address the cosmological constant problem (we will refer to such scalars as DE in the following). The presence of such extra scalar fields will manifest itself in additional attractive fifth forces, on which bounds can be placed from terrestrial and astronomical measurements. The juxtaposition of the dark energy and the absence of evidence for fifth forces are addressed by a range of screening mechanisms (see Refs.~\cite{Joyce:2014kja,Slosar:2019flp,Brax:2021wcv} for a review) that aim to naturally suppress the DE scalar's dynamics in laboratory environments (terrestrial or planetary) where measurements should be readily available to constrain them. Put differently, an environmentally aware DE scalar field that responds either through its mass or inertia in relation to its surroundings can naturally avoid fifth force constraints.

Recently in Ref.~\cite{Burrage:2018dvt}, it was shown that wide classes of scalar-tensor theories can be understood as Higgs mixing models to leading approximation. This motivates the phenomenology of the Higgs boson and its modifications compared to the SM from a cosmological perspective, elevating it beyond its importance for the electroweak scale and its instability in the SM. Dynamics observed at, e.g., the LHC can therefore bear relevance for cosmological scenarios and model building. That said, searches for DE-inspired interactions are not new~\cite{Brax:2009aw,Brax:2009ey,Brax:2015hma,Brax:2016did} and associated re-interpretations of standard BSM searches have been presented by, e.g., the ATLAS experiment targeting missing energy in Ref.~\cite{ATLAS:2019wdu}. 

Such final states can also be motivated by dark sector interactions and any associated discovery at the LHC would therefore not necessarily point out cosmologically relevant implications. In this work we show that the environmental sensitivity of DE scalars, however, gives rise to a new class of phenomenological signatures that is currently not investigated. In particular, these signatures predict characteristic correlations in the comparison of the ATLAS and CMS sensitivities as a result of their different detector designs. When their data is analysed {\emph{jointly}}, they can in principle discriminate cosmologically relevant screening mechanisms from other, competing scenarios such as hidden valleys~\cite{Han:2007ae}. While the parameter region that can be accessed this way is model-dependent and might not directly provide superior bounds when viewed in comparison against fifth force \cite{Upadhye:2012qu,Jaffe:2016fsh,Sabulsky:2018jma}, and other high precision \cite{Brax:2021owd,Brax:2022olf} measurements, these scenarios give rise to a novel class of signature-driven LHC analyses that primarily monetizes the complementary detector design that the different experiments offer. Open questions remain in the context of dark energy models in how far the semi-classical arguments employed in our work generalise to a full quantum description of the dark energy scalar at production. (A qualitative discussion is given after the LHC sensitivity region is identified.)

\section{Screened Scalar Fields}
\label{sec:envsen}
\begin{figure}[t]
\centering
\includegraphics[width=0.46\textwidth]{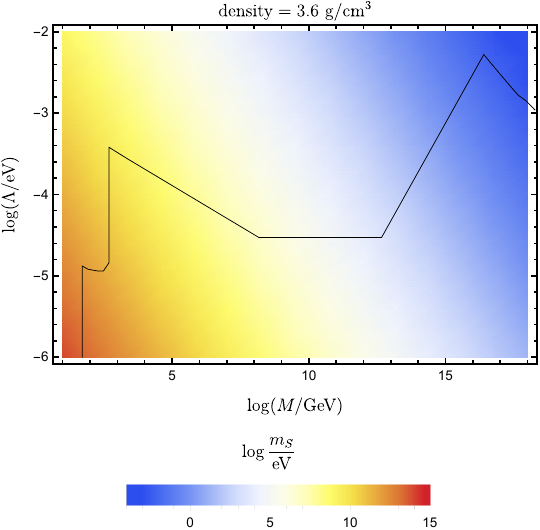}
\caption{The mass of the chameleon scalar, with a $\Lambda^5/\phi$ potential, as a function of $M$ (horizontal axis) and $\Lambda$ (vertical axis) when the density of the environment is $3.4 \mbox{ g/cm}^3$. Regions above the black line (roughly) are excluded by the constraints of previous experiments \cite{Brax:2021wcv,Brax:2022olf}, including atom-interferometry and E\"{o}t-Wash searches for fifth forces. Couplings below $M= 1~\text{GeV}$ are excluded by measurements of EW precision observables \cite{Brax:2009aw}. }\label{fig:chamminus_mass}
\vspace{-0.4cm}
\end{figure}

We focus on an illustrative model in which the scalar $s$ couples to a toy Standard Model through the Higgs portal. Our simplified Standard Model has a Higgs field $H$ and a fermion $\psi$ described by the Lagrangian (already assuming unitary gauge)
\begin{multline}
\mathcal{L} \supset -\frac{1}{2}(\partial s)^2 -\frac{1}{2}(\partial H)^2  -V(s) +\frac{1}{2} \mu_h^2 H^2 -\frac{\lambda}{4\!}H^4 -A_{hs}(s)H^2  
-i \bar{\psi} \slashed{\partial} \psi -\lambda_{\psi}H \bar{\psi} \psi\label{eq:higgsportal} 
\end{multline}
where $\mu_h$ is the tachyonic mass of the Higgs, $\lambda$ the quartic Higgs self-coupling and $\lambda_{\psi}$ the Yukawa coupling of our fermion to the Higgs. The dimensionful function $A_{hs}(s)$  describes the Higgs portal couplings. Higgs portal couplings of this form are a motivated choice for dark matter and other beyond-the-SM scalars~\cite{Binoth:1996au,Schabinger:2005ei,Patt:2006fw,Englert:2011yb}. Scalars originating from theories of dark energy and modified gravity are more commonly considered to couple  to matter through a conformal rescaling of the metric, for reviews see \cite{Copeland:2006wr,Clifton:2011jh}.  In Ref.~\cite{Burrage:2018dvt} (see also \cite{Copeland:2021qby, SevillanoMunoz:2022tfb}) it was shown that this is equivalent to a Higgs portal coupling at leading order.
Variation of equation (\ref{eq:higgsportal}) gives the equations of motion for $H$ and $s$:
\begin{equation}
\begin{split}
    \Box H & +(\mu_h^2 -2 A_{hs}(s))H -\lambda H^3 -\lambda_{\psi}\bar{\psi} \psi =0 \,,\\
    \Box s & -V^{\prime}(s) +A^{\prime}_{hs}(s)H^2=0\,.
\end{split}
\end{equation}
As an example of a dark energy model with screening, we focus on the chameleon model \cite{Khoury:2003rn,Mota:2006ed}, for which we choose\footnote{The relationship between these functions is a result of the field redefinitions needed to put a theory with a conformal coupling into canonical form~\cite{Burrage:2018dvt}. }
\begin{subequations}
\label{eq:AVhs}
\begin{align}
    A_{hs}(s)& =-\frac{\Lambda^5}{M^2s}+\mu_h^2\left(\frac{s}{M} +\frac{s^2}{M^2}\right) \label{eq:Ahs}\\
    V(s)& = \frac{\Lambda^5}{s}-\frac{v^2\mu_h^2s}{M} \,.\label{eq:Vs}
\end{align}
\end{subequations}
where $\Lambda$ is a dimensionful constant which controls the self interactions of the scalar field. Current constraints on the parameters $\Lambda$ and $M$ are shown in Figure \ref{fig:chamminus_mass}.  Values of $\Lambda$ around the cosmological constant scale of $\Lambda \sim 10^{-3} \mbox{ eV}$ are particularly well suited for cosmology.  This motivates us to study relatively strong couplings, where $M \sim \mbox{TeV}$, where the current constraints on $\Lambda$ are weaker. Higher powers of $s/M$ may be present in these functions, but their suppression by at least $1/M^3$ means that we can neglect them in this analysis as we will focus on $M\gtrsim \text{TeV}$.

After electroweak symmetry breaking the interactions between the Higgs and the chameleon scalar are
 \begin{equation}
\mathcal{L} \supset 2 v m_{sh} sh + m_{sh}sh^2+2 v\lambda_{hs} s^2h +\lambda_{hs} s^2 h^2
\end{equation}
where
\begin{align}
m_{sh} &= \frac{\Lambda^5}{M^2v_s^2}+\frac{\mu_h^2}{M}\left(1+\frac{2v_s}{M}\right)\\
\lambda_{hs} &= \frac{1}{M^2}\left(\mu^2_h-m_{s, \rm{eff}}^2\right)
\end{align}
where $v_s$ is the expectation value of $s$ and $m_{s, \rm{eff}}$ is the effective mass of the scalar field, to be defined shortly. The mass matrix for the two scalars can be diagonalized, and the mixing angle between the two fields is 
\begin{equation}
    \sin\theta \equiv s_\theta \simeq \frac{2  v}{M}\,.
\end{equation}
to leading order in $1/M$. In screened scalar field theories the non-linearities of the equation of motion for $s$ mean that the properties of fluctuations in the field depend on the background field profile, which in turn depends on the experimental environment.  To see this in our model, we start by considering situations in which the effects of the Higgs field are well described by the classical field value, which satisfies
\begin{equation}
\lambda H^2 = \mu_h^2-2A_{hs}(s) -\frac{\lambda_{\psi}}{H}\bar{\psi}\psi\,.
\end{equation}

We now assume that we are working in an environment with a background density of fermions. When these fermions are non-relativistic, their energy-momentum tensor can be related directly to the mass term in their Lagrangian, such that we can write; $\rho_{\psi} = \lambda_{\psi} H_m \langle \bar{\psi} \psi\rangle$, where $H_m$ is the value of $H$ at the minimum of the potential (which may be close to but not at the vev $v$). This expression can be interpreted as a mean-field approximation for the non-relativistic limit of the fermion energy-momentum tensor, valid when taking the classical limit in the case of high occupation numbers. We now perform this mean-field expansion, under the assumption that the heavy Higgs field, $H$, is slowly varying in the background environment, and will consider the equation of motion for fluctuations of the light mode, s, to first order. This is to say that we perform both a zeroth-order semi-classical approximation and a zeroth-order gradient expansion. 

Assuming that the Higgs vev is not significantly perturbed from its Standard Model value  $\langle H\rangle \approx v = \mu_h/\sqrt{\lambda}$, we find that at first order in perturbations, the vev is 
\begin{equation}
    \langle H\rangle=v\left[1-\frac{\langle s \rangle}{M}-\frac{\langle s \rangle^2}{M^2}-\frac{1}{2\lambda}\left( \frac{\lambda_{\psi}\langle \bar{\psi}\psi\rangle}{v^3}-\frac{2\Lambda^5}{M^2 v^2\langle s \rangle }\right)\right]\,.
\end{equation}
We now substitute this into the equation of motion for $s$, to find an equation independent of $H$,  where $s$ is sourced by the fermion fields. We set the mass of the fermion to be $m_{\psi} = \lambda_{\psi} v$, and assume that around large bodies of dense matter, we can replace $\bar{\psi}\psi \rightarrow \langle \bar{\psi}\psi\rangle = n_{\psi}$ where $n_{\psi}$ is the fermion number density. 
Then, to first order in $1/M$ the equation of motion for the scalar becomes
\begin{equation}
    \label{eq:mass}
    \Box s = -\frac{\Lambda^5}{s^2}-\frac{\rho_{\psi}}{M}\,,
\end{equation}
where $\rho_{\psi}= m_{\psi}n_{\psi}$ is the fermion mass density, and we see that the scalar field $s$ moves in an effective potential 
\begin{equation}
V_{\rm eff}(s)=\frac{\Lambda^5}{s}+\frac{\rho_{\psi}s}{M}\,.
\label{eq:Veff}
\end{equation}
We choose this popular and widely studied form of the chameleon effective potential (following from~Eq.~\ref{eq:AVhs}), which we will employ in the following, to highlight the general phenomenological implications of such screening mechanisms.\footnote{Inverse potentials can be criticized. However, we note that, e.g., symmetron scenarios~\cite{Hinterbichler:2011ca} will lead to similar phenomenological implications, albeit less transparent in their treatment.} 

Of course, in macroscopic matter the mass and density comes not just from the density of fundamental fermions, and their masses, but also from the gluon content of nuclei, and the QCD binding energy of protons and neutrons. In the context of light scalar couplings to matter this has been discussed previously, for example in Ref.~\cite{Burrage:2018dvt}.  A coupling between the light scalar and gluons is mediated by the mixing with the Higgs and the conformal anomaly.  At low energies this results in an effective vertex coupling of the light field to two gluons, which results in an effective coupling to nucleons at low energies (analogously to the effective coupling of the Higgs to nucleons \cite{Shifman:1978zn,Jungman:1995df,Gunion:1989we}). From this effective coupling to nucleons we can make the same mean field approximation to the background density as discussed for fermions above.

The form of the effective potential in Eq.~\eqref{eq:Veff} directly implies that the mass of the chameleon scalar field becomes dependent on the local matter density. The scalar mass increases in regions of high density and decreases in regions of low density.  This suppresses the production and interactions of the scalar in dense regions, such as the early universe, the interior of supernovae or fifth force
measurements and constitutes the fundamental property of any general screening mechanism (including Vainshtein models~\cite{Vainshtein:1972sx,Babichev:2013usa}).
The effective mass of the chameleon scalar is (in natural units)
\begin{equation}
    m_{s, \rm{eff}}^2 = 0.21 \left(\frac{\mbox{TeV}}{M}\right)^{3\over2} \left(\frac{10^{-3}\mbox{ eV}}{\Lambda}\right)^{5\over 2} \left(\frac{\rho_{\psi}}{\mbox{gcm}^{-3}}\right)^{3\over 2} \mbox{ GeV}^2
    \label{eq:chammass}
\end{equation}
We can expect the scalar to be heavy inside high density detector layers; see Fig.~\ref{fig:chamminus_mass} for an example of how the mass in this model varies with the parameters. In this example,  the mass of the light scalar varies from $\sim 10^{-4}\mbox{ eV}$ to $\sim 10^{14}\mbox{ eV}$ over the parameter space we consider for a density $\rho \sim \mbox{g/cm}^3$, corresponding to Compton wavelengths ranging from $\sim0.1\mbox{ cm}$ to $\sim 10^{-17}\mbox{  m}$. Scalars with masses smaller than $\sim 10^3\mbox{ eV}$ have Compton wavelengths larger than $\sim10^{-10}\mbox{ m}$, a typical scale for inter-atomic spacing in ordered solids, and so such scalars see a smoothed density, averaged over their Compton wavelength, as assumed in eq.~(\ref{eq:chammass}).

Equation (\ref{eq:chammass}), also implies that in some parts of the parameter space the scalars may be so heavy that they could resolve atomic nuclei, in such cases a full solution to the background scalar field profile requires modelling the dense atomic nuclei and the diffuse medium in between. We know that in regions of nuclear density, the chameleon field could be extremely heavy; when the density is $~ 10^{14}\mbox{ g/cm}^3$, scalar masses could reach $10^{10}\mbox{ GeV}$ for our benchmark parameters. The value of the light field inside the nucleus is also extremely small. In the near vacuum of inter-atomic space the field can relax to larger field values rising to reach at most a value where its mass corresponds to a Compton wavelength of order the inter-atomic spacing  \cite{Brax:2013cfa,Burrage:2014oza}. So, whilst the background field is not simply constant in this example, on average it takes a value such that fluctuations in the field have a Compton wavelength of at least the inter-atomic spacing. The expression we give in equation (\ref{eq:chammass}) does not capture all of these details of the possible variation in mass due to atomic structure of the detector, nevertheless we believe it is a useful indicative example of the way the background field and the mass of fluctuations behaves, and provides a guide to the novel collider signatures that could arise. 

As the scalar moves from one detector region to another, we will assume that the particle's mass changes instantaneously at the boundary and impose conservation of momentum at this point. The scalar particles we consider are relativistic, and their masses are such that their Compton wavelengths are always short, $\lesssim \mbox{m}$,  not only in the detector but also in the surrounding atmosphere.  Therefore, for the parameters we consider, we can assume that the background scalar field profile has adapted to the density profile of the detector, smoothed on the scale of the Compton wavelength. The field will evolve to smaller values in regions of higher density and larger values in regions of smaller density.

We assume that the scalar particles are created, in a particle collision, in the high quality vacuum of the beam pipe (We will give a quantitative estimate of the associated limitations further below).  We do not take into account here how the quark-gluon field of the colliding protons might influence the creation of such a scalar field.  This challenging analysis is left for future work. Such scalar particles with an environmentally dependent mass have also been searched for by the CAST experiment \cite{CAST:2015npk}, although relying on different signatures than the ones proposed here. Hence, we will assume all interactions to be point-like following the standard parton model. This treatment takes inspiration from studies of colour reconnection~\cite{Christiansen:2015yca}. For instance, the Higgs boson is known to be insensitive to hadronisation effects of the remaining event although it can interact with the beam remnants through effective interactions that are exploited in, e.g. dark matter searches. This is due to the Higgs boson's large lifetime, which is mirrored to an even larger extent for the dark energy scalar we consider.

The high-energy beam environment can, in principle, significantly modify this assumption. The creation of the scalar degree of freedom could proceed out-of-equilibrium, which is qualitatively different from other astronomical sources of chameleon production. The mass of the scalar at production for our semi-classical consideration is a crucial parameter as this determines the “active” area sampled by the dynamical mass via the Compton wave length as a characteristic parameter. Such out-of-equilibrium considerations present an important caveat, and our results hinge on a better understanding of chameleons in such an environment. Concretely, even the semi-classical shape of the chameleon potential in a relativistic non-homogenous medium is not known; higher order effects are likely to play a fundamental role. Limits on the Compton wavelength will set a lower bound on the scalar mass, and we will analyse the scalar field equation in the energetic oscillating background to provide a quantitative estimate once we have identified the relevant parameter region for the LHC experiments in the next section.

We also stress that in our model, the Higgs can decay into pairs of dark energy scalars (such interactions arise, e.g., from the Higgs potential in the mass basis) and this process is expected to avoid beam pollution given the results related to colour reconnection~\cite{Christiansen:2015yca}, where it is understood that the Higgs decays essentially in vacuum. Therefore, the limits on the energy scale $M$ derived from the Higgs measurements presented below, are independent of the beam environment. This avenue could be further explored in the future.

\section{Displaced signs of Screening}
\label{sec:displ}
The most stringent constraints on the scenario described above in a collider environment are from invisible Higgs decays which bound $\lambda_{hss}$ and the mixing angle that modifies the visible Higgs signal strengths $\mu\simeq c^2_\theta$~\cite{Englert:2011yb,Bauer:2020nld}. The extra scalar particle inherits its visible phenomenology from that of the Higgs boson as a function of the mixing. Its decay width then becomes a function of its environmentally dependent mass. In particular, if the scalar is lighter than the two-pion or lepton thresholds, only loop-induced decays into photons remain as efficient decay channels. Numerical investigation shows that for such parameter regions, the scalar becomes stable on collider length scales, effectively removing its detectability at the LHC given the latter's trigger criteria.\footnote{A significant production via missing energy signatures would require large mixing angles which are constrained by current Higgs measurements.} In the following we will employ a full calculation of $s\to$~SM final states at 1-loop level, which we have validated against publicly available tools such as {\sc{Hdecay}}~\cite{Djouadi:2018xqq}.

\begin{table*}[t]
\centering 
\begin{tabular}{|c|c|c|c|c|c|c|}
\hline
\multirow{2}{*}{ } & \multicolumn{3}{c|}{ATLAS} & \multicolumn{3}{c|}{CMS} \\ \cline{2-7}
			 & $R_{\mathrm{in}}$ [mm] & $R_{\mathrm{out}}$ [mm] & $\rho$ [g/cm$^3$] & $R_{\mathrm{in}}$ [mm] & $R_{\mathrm{out}}$ [mm] & $\rho$ [g/cm$^3$] \\ \hline
Beam pipe & 23.5 & 24.3 & 1.85 & 23.5 & 24.3 & 1.85 \\ \hline
Tracker & 33.25 & 1082 & 0.12 & 29 & 1233 & 0.011 \\ \hline
EM Calor. & 1500 & 1970 & 3.6 & 1238 & 1750 & 8.3 \\ \hline
Had. Calor. & 2280 & 3920 & 7.5 & 1770 & 2950 & 8.3 \\ \hline
Solenoid & 1150 & 1390 & 1.4 & 2950 & 4020 & 0.3 \\ \hline
Muon spectr. & 4415 & 10675 & 0.003 & 4020 & 7380 & 3.7 \\ \hline 
\end{tabular}
\caption{Inner and outer radii and densities of the different layers of the simplified ATLAS and CMS model. We model the detector as layers of uniform density; the geometry towards the endcaps is not relevant for the kinematics of the process that we study in this work as the scalar $s$ is centrally produced.}\label{tab:detectors}
\end{table*}

Phenomenologically more interesting are signatures featuring displaced decays. The decay of a sample of scalar particles as a function of the particles' speed $\beta$ (in natural units), their decay width $\Gamma$ and length $\ell$ is given by
\begin{equation*}
N(\ell,\beta,\Gamma) =N_0 \exp\left( - { \sqrt{1-\beta^2} \ell \,\Gamma\over \beta}\right)\,.
\end{equation*}
The decay width will depend on the particles' mass, which depends on the detector geometry as provided in Tab.~\ref{tab:detectors} through the dependence $m_{s, \rm{eff}}=m_{s, \rm{eff}}(\rho)=m_{s, \rm{eff}}(\ell)$. Continuity of the decay function across adjacent detector layers can then be used to construct a decay probability distribution which depends on $\beta^2 \simeq p_T^2 ( m^2(\ell) + p_T^2)^{-1}$ with $p_T$ as the transverse momentum. 

For the detector description, a simplified model is used based on the geometry of the ATLAS~\cite{ATLAS:2008xda} and CMS~\cite{CMS:2008xjf} detectors. This assumes that the detectors are composed of cylindrical shells of uniform density. The density is calculated approximately, taking into account the different materials and layout of each sub-detector, which should at least qualitatively reproduce the main differences between the two detectors, and in particular the much denser electromagnetic calorimeter of CMS, which drives the difference in the decays of the scalar discussed below. 
Deviations from a realistic simulation are expected, especially for sub-detectors composed of various parts of significantly different densities, such as the trackers, sampling calorimeters and the muon spectrometers. A realistic simulation of the signals would need to include the $\rho$-dependence of the particles' lifetime in the \textsc{Geant4} toolkit~\cite{GEANT4:2002zbu}, which is left for future work.

Sizeable production of the DE scalar is challenging for parameter regions close to $\cos^2\theta\simeq 1$ identified by recent LHC Higgs measurements. To overcome the
mixing angle suppression a significant SM-like production cross section is required. This is provided through associated production $pp \to s Z$ with decay $Z\to~\text{light leptons}~(l=e,\mu)$, which is large for $m_s\lesssim 1~\text{GeV}$ at 13 TeV LHC collisions with a cross section of 
\begin{equation}
\sigma(pp \to sZ_l) = 15.7~\text{pb} \times s^2_\theta\,.
\end{equation}
This includes next-to-leading order QCD corrections obtained with \textsc{MadGraph}\_aMC@NLO~\cite{Alwall:2014hca}; the transverse momentum distribution of the scalar $s$ peaks sharply with expectation value $\langle p_T(s)\rangle = 8.5~\text{GeV}$ and the scalar is produced in the central part of the detector $|\eta_s| < 2.5$, thus rendering all layers of the detectors relevant for our analysis. Given the relatively good localisation, we do not convolve with the $p_T,\eta$ probability distributions in the following (which will slightly wash out our findings below). This final state resembles Drell-Yan $Z$ production and can therefore be triggered at the LHC using the prompt leptons from the $Z$ decay~\cite{ATLAS:2021jig}. 

\begin{figure}[t]
    \centering
    \begin{subfigure}
        \centering
         \includegraphics[width=0.46\textwidth]{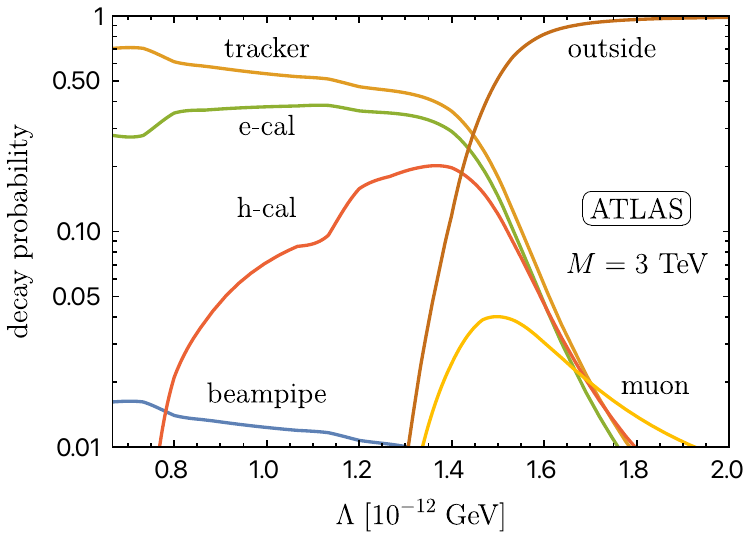}
            \end{subfigure}
    \hfill
    \begin{subfigure}
        \centering
         \includegraphics[width=0.46\textwidth]{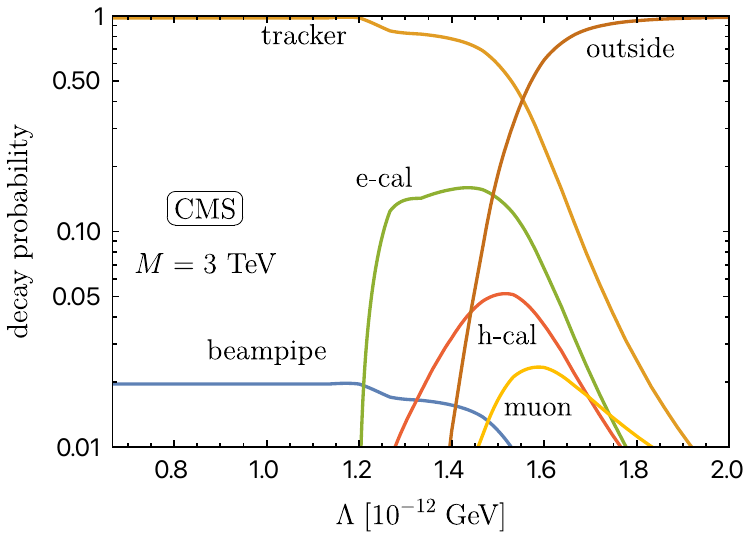}
            \end{subfigure}
    \caption{ Representative decay probabilities for $M=3~\text{TeV}$ (thus avoiding Higgs signal strength constraints) as a function of $\Lambda$, comparing ATLAS (left) and CMS (right) for $p_T(S)\simeq 8.5~\text{GeV}$.}
    \label{fig:m3tev}
\end{figure}

Given the typical mass scales that are determined by the densities of Tab.~\ref{tab:detectors}, this means that we have $\beta \simeq 1$ and the decay in this scenario is determined predominantly by the environmental response of the mass for given choices of $\Lambda,M$. The latter parameter is constrained by the constraints on the Higgs mixing in presence of additional invisible Higgs decay channels 
\begin{equation}
\mu = {\left[\sigma(h)\text{BR} \right]^\text{BSM} \over \left[\sigma(h)  \text{BR} \right]^\text{SM}}
= c^2_\theta \left(1+ {\Gamma(h\to ss)\over \Gamma^\text{SM}(h\to \text{all}) \, c^2_\theta }\right)^{-1}.
\end{equation}
As $s_\theta,\lambda_{hss} \sim M^{-1},M^{-2}$, a choice of $M\gtrsim 3~\text{TeV}$ avoids Higgs constraints $\delta \mu/\mu \lesssim 10\%$ from current data, see Ref.~\cite{ATLAS:2023tkt,CMS:2023sdw}. For this parameter choice we plot the decay probability in Fig.~\ref{fig:m3tev} for a scalar produced with $p_T\simeq 8.5~\text{GeV}$. For the mixing value that results from these parameter choices we still have a sizeable production cross section of the order of 0.5~\text{pb} which could imply detectable signatures also for low decay probabilities depending on the detector performance. Active final states are typically light leptons, photons and pions that are active for effective masses of $m_S\simeq 1~\text{GeV}$ and we can therefore expect possibly good detection prospects in the tracker, electromagnetic, and hadronic calorimeters.

\begin{figure}[t]
\centering
\includegraphics[width=0.46\textwidth]{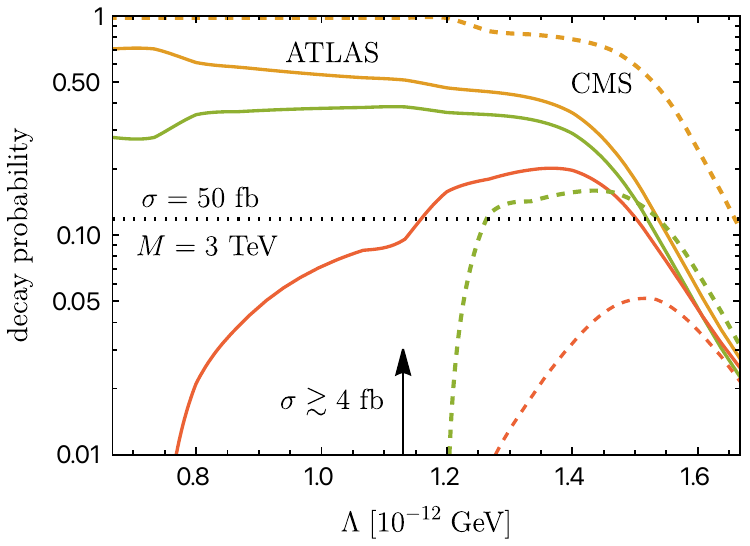}
\caption{\label{fig:m3tev2} Comparison of different detector regions of ATLAS (solid) and CMS (dashed) responding to the same parameter points as in Fig.~\ref{fig:m3tev}. Colours are identical to Fig.~\ref{fig:m3tev}. A $1\%$ decay probability corresponds to a cross section of around 4~fb, the dashed lines show the equivalent of a $50~\text{fb}$ cross section for comparison. 
}
\vspace{-0.4cm}
\end{figure}

The environmental response of the scalar boson then opens up a phenomenological possibility that is vastly different from hidden valley scenarios (discussed in Refs.~\cite{Han:2007ae,Baumgart:2009tn,Alimena:2019zri}). In the latter case, we might be dealing with complicated decay chains in the hidden sector, however, the scalar's lifetime is a function of the underlying theories' parameters; decay lengths are identical for production at ATLAS and CMS. As a consequence, the expected phenomenology is highly correlated between ATLAS and CMS as such states will be produced through identical production processes at either experiment. For the case discussed above, this is dramatically different. While significant production modes are identical the different detector geometries lead to different ``active'' decay areas when comparing the data sets obtained at ATLAS and CMS. The major difference between the two detectors is related to their electromagnetic calorimeter design. CMS is comparably thin and dense which leads to a very acute stopping power in this particular layer for a particular mass range. ATLAS in comparison is much wider which enables significant decay while traversing that layer. This extends to the comparison of the hadronic calorimeters as well, for which ATLAS is slightly less dense, yet considerably wider. This qualitative behaviour also becomes apparent in the comparison given in Fig.~\ref{fig:m3tev2}: In the region where either experiment will see displaced vertex signatures with significant cross sections, ATLAS and CMS could observe such states with identical probabilities when comparing the hadronic (ATLAS) and electromagnetic (CMS) calorimeter information. In case a signature is found such a dependence acts as a direct discriminator between the scenarios discussed in this work and hidden valley models. In comparison to Fig.~\ref{fig:chamminus_mass} we can therefore see that the LHC can gain sensitivity in a small, yet relevant parameter region of our Chameleon example that complements existing searches, specifically employing the complementarity of ATLAS and CMS detectors. 
%

We stress that these signatures are not directly comparable to existing searches without making particular assumptions related to the detector performance. Focusing on chameleon decays in the ECal as an illustrative example, we find that chameleon decays are triggered when the mass exceeds the hadronic threshold $s\to \pi^0 \pi^0$ followed by rapid decays of the $\pi^0$ into photons. The experimental signature that would result from the $pp \to s Z$ process considered above, would therefore consist of two prompt leptons and two displaced neutral pions. In the light of searches involving similar signatures~\cite{CMS:2023tie,ATLAS:2023meo}, we can anticipate a total acceptance times efficiency of the order of 0.1-1\%. This representative acceptance still corresponds to $\sim $120-1200 signal events at each experiment with rates correlated according to Fig.~\ref{fig:m3tev2}. Whilst a correlation across different detector regions would be ideal, such an observation would be a strong indication of the underlying theory presented here already.

It is now a good point to return to this model's initially pointed out caveat, namely the environmental dependence at production. The Chameleon, through Eq.~\eqref{eq:mass}, in the LHC production environment depends on $\Lambda\simeq 10^{-12}\mbox{ GeV}$, the environmental energy density $\rho_\psi$ and an (effective) scalar mass at production that can be readily added to Eq.~\eqref{eq:mass}. The LHC collides proton bunches with a characteristic frequency of 40~MHz, providing the periodicity of the $\rho_\psi$ which needs to be contrasted with the field's oscillation through its Compton wavelength at production. We can assume $\rho_\psi/M \sim \text{TeV}^3$ and the LHC bunch crossing induces a mass scale of $m_b\simeq 10^{-17}~\text{GeV}$. We employ the Runge-Kutta 4 method to qualitatively analyse the solution in this environment in comparison with the free scalar field that is implicitly assumed in our perturbative approach. For the time-dependencies of the Chameleon field for these parameters, we find that for $m_s\gg 10^{-17}~\text{GeV}$ is largely unperturbed by the comparably slow oscillation of the external source. The solutions resemble those of the free Klein-Gordon field, $\rho_\psi$ determines an adiabatic offset of the wave solution (which can be removed through an appropriate field redefinition). The spatial dependence along the beampipe can again be approximated by an oscillation with the characteristic bunch crossing (i.e. it can be obtained from the time-evolution via a Wick rotation). Again we find good agreement between free and environment-active cases for $m_s\gg 10^{-17}~\text{GeV}$. On the contrary when $m_s\lesssim m_b$, we observe parametric resonance and an increased amount of anharmonicity. Our phenomenological discussion above is likely to be inaccurate as quantisation needs to take increasingly respect that backreaction with the oscillating background.

Notwithstanding the aforementioned out-of-equilibrium caveat, we can be somewhat confident that the semi-classical scalar mass at production in the parton model is justified if it is heavy enough compared to the LHC's bunch crossing. However, if the scalar indeed traces, non-locally, beam energies on the characteristic time scales of particle creation beyond modifying the classical solution, the out-of-equilibrium production at colliders will resemble the high density, equilibrium conditions of astrophysical laboratories. The interpretability of the proposed signature along the lines of dark energy therefore depends on a better understanding of the non-local chameleon response to a non-homogenous relativistic beam medium, which is beyond the scope of this analysis. We note, however, that electron-positron collisions at past and future colliders could potentially mitigate such a caveat. In parallel, the decay $h\to ss$ will avoid the aforementioned beam-related caveats and therefore provides an alternative avenue for $s$ generation, which we leave for future work.  Nevertheless, in the screening models that motivate this study, the environmental correction to the light scalar mass depends on the trace of the energy-momentum tensor of the environment, $T= \rho - 3p$, which is well approximated by the energy density in a non-relativistic setting such as a detector but will approach zero in relativistic environments such as a particle beam.

\section{Summary}
\label{sec:sum}
Sensitivity to beyond the Standard Model interactions with direct relevance to astrophysical phenomena such as dark energy might manifest themselves in displaced vertex signatures at the Large Hadron Collider. While such analyses are motivated from a range of new physics models, their behaviour in relation to astrophysically relevant screening mechanisms will only become clear when the differences in the ATLAS and CMS detector design are considered. These differences then become vital in understanding the nature of an associated discovery with potential paradigm-shifting consequences for our understanding of cosmological evolution. We have shown that cosmologically-relevant scalars can be sizeably produced at the LHC, while further decaying with probabilities correlated with the detector design. Our findings suggest that the LHC experiments could indeed possess significant sensitivity to such scenarios which could act as a major discrimination tool when a displaced vertex BSM discovery is made. Our sensitivity analysis might not directly translate to more realistic sensitivity estimates when detector effects etc. are included, but we have estimated that residual sensitivity should remain even for pessimistic efficiencies. Further work is required to stress-test the semi classical validity of the parton model in the context of dark energy models including the Chameleon.

\medskip
\subsection*{Acknowledgements}We would like to thank Martin Bauer for useful conversations in the early stages of this work. S.A. is supported by the German Research Foundation (DFG) under grant No. AR 1321/1-1.
C.B. is supported by a Research Leadership Award from the Leverhulme Trust and by the STFC under grant ST/T000732/1.
C.E. is supported by the STFC under grant ST/T000945/1, the Leverhulme Trust under grant RPG-2021-03, and the IPPP Associateship Scheme. 


\bibliography{paper_final}
\end{document}